\documentstyle[12pt]{article}
        \catcode`\@=11
        \@addtoreset{equation}{section}

\begin{document}
\makeatletter
\def\siml{\mathrel{\mathpalette\gl@align<}}
\def\simg{\mathrel{\mathpalette\gl@align>}}
\def\gl@align#1#2{\lower.6ex\vbox{\baselineskip\z@skip\lineskip\z@
 \ialign{$\m@th#1\hfill##\hfil$\crcr#2\crcr{\sim}\crcr}}}
\makeatother
\hbadness=10000
\hbadness=10000
\begin{titlepage}
\nopagebreak
\def\thefootnote{\fnsymbol{footnote}}
\begin{flushright}

        {\normalsize
 LMU-TPW 95-5\\
April, 1995   }\\
\end{flushright}
\vspace{1cm}
\begin{center}
\renewcommand{\thefootnote}{\fnsymbol{footnote}}
{\large \bf Selection Rules for Nonrenormalizable Couplings
in Superstring Theories}

\vspace{1cm}

{\bf
Tatsuo Kobayashi
\footnote[1]{Alexander von Humboldt Fellow \\
\phantom{xxx}e-mail: kobayash@lswes8.ls-wess.physik.uni-muenchen.de}
}

\vspace{1cm}
       Sektion Physik, Universit\"at M\"unchen, \\

       Theresienstr. 37, D-80333 M\"unchen, Germany \\

\end{center}
\vspace{1cm}

\nopagebreak
\begin{abstract}
We study nonrenormalizable coupling terms in $Z_N$ orbifold models.
Nontrivial selection rules of couplings are provided and
 cannot be understood in terms of a simple symmetry of effective
field theories.
We also discuss phenomenological implications of theses selection rules
for the quark mass matrices.

\end{abstract}

\vfill
\end{titlepage}
\pagestyle{plain}
\newpage
\voffset = 0.5 cm

\vspace{0.8 cm}
{\large \bf 1. Introduction}
\vspace{0.8 cm}

Superstring theory is the only known candidate for the unified theory of all
the interactions including gravity.
Much work has been devoted in order to study phenomenological aspects of
effective field theories derived from superstring theory.
Nonrenormalizable couplings lead to renormalizable coupling terms with
suppression factors after symmetries are broken like an anomalous U(1)
symmetry breaking \cite{anom,anom1,anom2}.
Thus nonrenormalizable couplings have been discussed in order to derive
some types of hierarchical structures, e.g., mass matrices of quarks and
leptons \cite{mass,IG}, the $\mu$-term \cite{mu}, large masses of
right-handed neutrinos \cite{neutrino} and so on.
In ref.\cite{IG} it is shown that one can obtain the realistic quark and
lepton masses and mixing angles within the framework of a simple extension
of the standard model by U(1), using nonrenormalizable couplings.
Hence it is important to investigate selection rules of nonrenormalizable
couplings as well as renormalizable couplings in superstring theory.

Coupling terms are restricted by several symmetries of superstring theory.
For orbifold models \cite{Orbi}, selection rules of Yukawa couplings are
discussed in refs.[9-13] and further selection rules of nonrenormalizable
couplings are discussed in refs.\cite{Cvetic,anom1,Nonreno}.
In ref.\cite{Nonreno} a nontrivial selection rule of nonrenormalizable
couplings is derived in the case where states sit at the
same fixed point of the orbifold.
One cannot understand this rule in terms of symmetries of effective fields
theories.
In this paper we study nonrenormalizable couplings in $Z_N$ orbifold models
to derive other nontrivial selection rules.
We also discuss phenomenological implications of theses selection rules.

This paper is organized as follows.
In section 2 we study selection rules of nonrenormalizable couplings due to
symmetries of a six-dimensional compactified space and the $H$-momentum
conservation.
In section 3 we discuss phenomenological implications of these selection
rules, e.g. for the quark mass matrices.
We also comment on the CP phase.
The last section is devoted to conclusion.

\vspace{0.8 cm}
{\large \bf 2. Nonrenormalizable couplings in orbifold models}
\vspace{0.8 cm}

In orbifold models, string states consist of the bosonic strings on the
four-dimensional space-time and a six-dimensional orbifold, their
right-moving superpartners and left-moving gauge parts.
The right-moving fermionic parts are bosonized and momenta of bosonized
fields, $H_t$ $(t=1 \sim 5)$, span an SO(10) lattice.
A $Z_N$ orbifold is obtained through a division of a six-dimensional space
R$^6$ by a six-dimensional lattice and its automorphism $\theta$.
We denote eigenvalues of $\theta$ in a complex basis $(X_i,\overline X_i)$
$(i=1 \sim 3)$ as exp$[2\pi i v^i]$.
For simplicity, we restrict ourselves to the $Z_N$ orbifolds which are
obtained by products of three two-dimensional orbifolds, i.e. $Z_3$,
$Z_4$, $Z_6$-I and $Z_6$-II orbifolds.
Their twists $\theta$ are shown in the second column of Table 1.
To preserve the world-sheet supersymmetry, the SO(10) lattice is also
divided by the shift $v^t$, whose fourth and fifth elements correspond to
the four-dimensional space-time and vanish.

There are two types of closed strings on the orbifolds.
One is the untwisted string and the other is the twisted string.
For the $\theta^k$-twisted sector $T_k$, the string coordinate
$x_\nu$ $(\nu=1 \sim 6)$ has the following boundary condition:
$$
 x_\nu(\sigma=2 \pi)=\theta^kx_\nu(\sigma=0)+e_\nu,
\eqno(2.1)$$
where $e_\nu$ is a lattice vector.
A zero-mode of the twisted string satisfy the same equation as (2.1) and
is called a fixed point.
The fixed point is denoted by the space group element $(\theta^k,e_\nu)$.
Note that fixed points of $\theta^k$ are not always fixed under $\theta$.
Therefore we have to take linear combinations of twisted states
corresponding to fixed points $f$ in order to provide eigenstates under
$\theta$.
Suppose that $\sigma_f$ is a $\theta^k$-twist field \cite{DFMS}
corresponding to $f$, which is fixed under $\theta^m$ with the minimum
number $m$.
Eigenstates of $\theta$ are obtained by linear combinations as \cite{KO1,KO2}
$$
\sigma_{f\gamma} \equiv \sigma_f+\gamma^{-1}\sigma_{\theta f}
+\gamma^{-2}\sigma_{\theta^2 f} + \cdots +
\gamma^{1-m}\sigma_{\theta^{m-1}f},
\eqno(2.2)$$
where $\gamma=\exp [2 \pi i n/m]$ and it is an eigenvalue of the $Z_N$
twist.
The $\theta^k$-twisted sector has the $H$-momentum $\tilde p^t=p^t+kv^t$,
where $p^t$ is a quantized momentum on the SO(10) lattice. The SO(10)
vector and spinor correspond to bosonic and fermionic states in the
space-time, respectively.
For each twisted sector, the $H$-momentum of bosonic massless states is
shown in Table 1 \cite{KO1,KO2}.
The massless bosons in the untwisted sector $U_j$ have the $H$-momenta
$p^i=\delta^i_j$ $(i,j=1 \sim 3)$.
The $H$-momenta of massless fermions are obtained from those of the
corresponding bosons by the space-time supertransformation \cite{FMS}.
The space-time supercharge includes the $H$-momentum $(-1,-1,-1)/2$ for
the six-dimensional internal space.

For bosons in the non-oscillated $T_k$ sector, right-moving parts of vertex
operators are obtained in the $-1$-picture as
$$
V_{-1}=e^{-\phi}e^{iKx}e^{i\tilde p H} \sigma_{f\gamma},
\eqno(2.3)$$
where $\phi$ corresponds to the bosonized superconformal ghost \cite{FMS} and
$e^{iKx}$ is the four-dimensional space-time part.
Right-moving parts of vertex operators for fermionic states are
written in the $-1/2$-picture as
$$
V_{-1/2}=e^{-1/2\phi}e^{iKx}e^{i\tilde p H} \sigma_{f\gamma}.
\eqno(2.4)$$
Similarly we can obtain vertex operators in the untwisted sector, where we
do not need twist fields.
We can change the picture by the picture changing operator \cite{FMS}, which
includes the following term:
$$
e^\phi e^{-i\alpha^iH}\partial X_i,
\eqno(2.5)$$
where $\alpha^1=(1,0,0)$, $\alpha^2=(0,1,0)$ and $\alpha^3=(0,0,1)$.
Note that vertex operators corresponding to the non-oscillated states
include oscillators $\partial X_i$ if we change their pictures.

Now we study on nonrenormalizable couplings of the type $V_FV_FV_B^\ell$,
where $V_F$ and $V_B$ are the vertex operators for fermions and bosons,
respectively.
The vertex operators consist of several parts.
Each part provides a selection rule due to charge conservation.
At first we discuss the selection rule due to the space group invariance.
A coupling vanishes unless a product of space group elements corresponding
to states includes the identity up to conjugacy classes.
Among the space group invariance, the point group invariance provides
an easy selection rule.
For 3-point couplings of the $Z_N$ orbifold models, the selection rules
due to the space group are explicitly obtained in ref.\cite{KO2,CM}.
We can extend those results to nonrenormalizable couplings.
Further a product of eigenvalues, $\prod_i \gamma_i$, should be the identity.

Next we discuss the $\phi$-charge conservation.
To match the background $\phi$-charge, the sum of the $\phi$-charges
of vertex operators should be $-2$ \cite{FMS}.
Thus we have to use the picture changing operator ($\ell-1$) times in the
correlation function $<V_FV_FV_B^\ell>$.
That causes change of $H$-momenta and appearance of oscillators $\partial X_i$.
The $H$-momenta should be conserved in nonvanishing couplings.
Further nonvanishing couplings should be invariant under the $Z_N$
rotation of oscillators $\partial X_i$ for each complex plane.
Note that oscillators of each $T_k$ sector, $\partial X_{i(k)}$, transform
under the $Z_N$ rotation in a different way from oscillators of other
sectors as follows,
$$
\partial X_{i(k)} \rightarrow e^{2 \pi i kv^i} \partial X_{i(k)}.
\eqno(2.6)$$

We discuss on coupling conditions of non-oscillated twisted sectors.
For a concrete example, we consider the $Z_4$ orbifold models.
There are two twisted sectors, $T_1$ and $T_2$ for matter fields.
The other twisted sector $T_3$ corresponds to anti-matter fields.
Massless states of $T_1$ and $T_2$ have the $H$-momenta
$\tilde p^i=(1,1,2)/4$ and $\tilde p^i=(2,2,0)/4$, respectively,
as shown in Table 1.
The point group invariance allows $T_1^{4\ell}$-couplings,
$(T_1^2T_2)^\ell$-couplings, $T_2^{2\ell}$-couplings and their products.
Here we investigate the $T_1^{4\ell}$-coupling.
The corresponding correlation function
$<V_{-1/2}V_{-1/2}V_{-1}^{4\ell-2}>$ has totally the $H$-momentum as
$(\ell-1,\ell-1,2\ell-1)$, where $-1$ is the contribution due to the
space-time supercharge in $V_{-1/2}$.
We use the picture changing operator $(4\ell-3)$ times in order to
change the total picture of the correlation function into $-2$.
At the same time we have to make the $H$-momentum conserved by using
(2.5).
As a result, the $H$-momentum conserved correlation function includes
oscillators of $T_1$ as
$$
(\partial X_{1(1)})^{\ell-1}
(\partial X_{2(1)})^{\ell-1} (\partial X_{3(1)})^{2\ell-1}.
\eqno(2.7)$$
Nonvanishing correlation functions should be invariant under the $Z_4$
rotation (2.6) of each two-dimensional plane.
This invariance requires $\ell-1=4m$ and $2\ell-1=2n$.
The latter is impossible to satisfy.
Thus the $T_1^{4\ell}$-couplings are not allowed.
Similarly the $T_2^{2\ell}$-couplings are forbidden.

Next we study $(T_1^2T_2)^{\ell+1}$-couplings.
Corresponding correlation functions with the total $-2$-picture include
oscillators in the $H$-momentum conserved form as follows
$$
(\partial X_{1})^{\ell}
(\partial X_{2})^{\ell} (\partial X_{3})^{\ell}.
\eqno(2.8)$$
Note that these oscillators include two types, i.e. oscillators of $T_1$
and $T_2$ and these oscillators $\partial X_{i(k)}$ transform under the $Z_4$
rotation differently from each other (2.6).
For the first and second planes, the invariance under the $Z_4$ rotation
requires that $\ell$ should be even.
In the case with $\ell=4m$, we can take the following combination of
oscillators:
$$
(\partial X_{1(1)})^{\ell}
(\partial X_{2(1)})^{\ell} (\partial X_{3(2)})^{\ell}.
\eqno(2.9)$$
This combination is invariant under the $Z_4$ rotation of each two-dimensional
plane (2.6).
In the case with $\ell=4m+2$ and $m>0$, we can take similarly the following
$Z_4$ invariant combination of oscillators:
$$
(\partial X_{1(1)})^{\ell-2} (\partial X_{1(2)})^2
(\partial X_{2(1)})^{\ell-2} (\partial X_{2(2)})^2
(\partial X_{3(1)})^4        (\partial X_{3(2)})^{\ell-4}.
\eqno(2.10)$$
For $\ell=2$ we can take the $Z_4$ invariant combination of oscillators as
follows,
$$(\partial X_{1(2)})^2(\partial X_{2(2)})^2 (\partial X_{3(1)})^2.
\eqno(2.11)$$
Note that this combination does not correspond to the \lq \lq standard" form
\break $<V_{-1/2}V_{-1/2}V_{-1}V_0^6>$.
If the two fermions belong to $T_1$, the combination (2.11) is obtained as
$<V_{-1/2}V_{-1/2}V_{-1}V_{-1}V_1V_0^4>$, where $V_1$ corresponds to $T_2$.
As a result, the $(T_1^2T_2)^{m+1}$-couplings are allowed if $m$ is even.

Similarly, we can obtain selection rules for
$(T_1T_1T_2)^{\ell +1}T_1^{4m}$-couplings and
$(T_1T_1T_2)^{\ell +1}T_2^{2m}$-couplings.
Both types are allowed if $\ell+m$ is even.
The $H$-momentum conservation and $Z_4$ invariance of oscillators require
nontrivial forms of correlation functions for smaller values of
$\ell$ and $m$.
Note that $T_1^{4\ell}T_2^{2m}(T_1T_1T_2)^{n}$-couplings correspond to either
of the above two types.

For the other orbifold models, we can derive allowed nonrenormalizable
couplings in a similar way.
The results are shown in Table 2, where PGI denotes the selection rule due
to the point group invariance, i.e. $\ell +2m+3n=6p$ for the
$T_1^\ell T_2^mT_3^n$-couplings of $Z_6$-I, $\ell +m+2n=3p$ for
the $T_1^{2\ell} T_2^mT_4^n$-couplings of $Z_6$-II and
$\ell +2m+3n+4p=6q$ for the $T_1^{\ell} T_2^mT_3^nT_4^p$-couplings
of $Z_6$-II.
For the $Z_3$ orbifold models, the allowed couplings have been obtained
already in ref.\cite{anom1,Nonreno}.
This selection rule is not difficult to understand in terms of an
$R$-symmetry of effective field theories.
For the $Z_6$-I orbifold models, $T_3^\ell$-couplings are forbidden.
On the other hand, the $Z_6$-II orbifold models do not allow any couplings
including only one twisted sector, i.e. $T_k^\ell$-couplings $(k=1\sim 4)$.
Further $T_2^\ell T_4^m$-couplings are forbidden.
For some couplings, the $H$-momentum conservation and $Z_4$ invariance of
oscillators require nontrivial forms of correlation functions which include
vertex operators with higher pictures.

So far we have studied the selection rules of $<V_FV_FV_B^\ell>$ including
only the non-oscillated twisted sectors.
Similarly we can discuss the couplings including the untwisted sectors and
the oscillated sectors.
For the untwisted sector, the $Z_N$ rotation acts as (2.6) with $k=0$.
Thus the $H$-momentum conservation and the $Z_N$ invariance (2.6) do not
forbid the couplings $UT^\ell$ if $T^\ell$ is allowed.
The $H$-conservation rule is very severe for the couplings including only
bosons.
Every massless bosonic state has the total internal $H$-momentum as
$\tilde P=\sum_{i=1}^3 \tilde p^i=1$.
Thus the correlation function $<V_{-1}V_{-1}V_0\cdots V_0>$ has
the total internal $H$-momentum $\tilde P=2$.
Therefore any couplings of only the bosons are not allowed.
This rule implies that we cannot generate the $\mu$-term through this
type of the couplings and we might need a fermion condensation \cite{mu}
to generate the $\mu$-term by the nonrenormalizable couplings.

In this section we have discussed on the selection rules due to the
$H$-momentum conservation and the $Z_N$ invariance of oscillators.
A further nontrivial selection rule is required if all twisted states sit at
the same fixed point as shown in ref.\cite{Nonreno}.

\vspace{0.8 cm}
{\large \bf 3. Phenomenological implications of nonrenormalizable
couplings }
\vspace{0.8 cm}

In this section we discuss on phenomenological implications of the results
obtained in the previous section.
At first we study whether the selection rules of nonrenormalizable couplings
can be understood by a simple symmetry of effective field theories.
We take the selection rule for the nonrenormalizable couplings
in the $Z_4$ orbifold models as an example.
We consider a $Z_4$ discrete $R$-symmetry and assign the $R$-charge
$k/2$ to the $T_k$ sector.
In addition we assign the $R$-charge 1 to the fermionic coordinate of the
superspace.
We impose that the superpotential should have the $R$-charge
2 mod 4.
This $R$-symmetry requires that $\ell$ should be even for the
$(T_1^2T_2)^{\ell+1}$-coupling.
However, there is no reason to forbid the $T_1^{8\ell +4}$-couplings and
the $T_2^{4\ell +2}$-couplings.

After symmetries are broken like the anomalous U(1) symmetry breaking
\cite{anom,anom1,anom2} and fields develop vacuum expectation values (VEVs)
$v$, some nonrenormalizable couplings work as renormalizable coupling terms
including a suppression factor $\varepsilon=v/M$, where $M$ is the Planck
scale or the string scale.
Several types of nonrenormalizable couplings in Table 2 lead to strongly
suppressed couplings.
That fact is favorable to the large hierarchy like the $\mu$-term.

In ref.\cite{RRR}, left-right symmetric mass matrices with texture zeros are
discussed to derive the quark masses and mixing angles consistent with the
experimental results.
Five solutions with five texture zeros are obtained.
In ref.\cite{IG} the possibility of deriving such mass matrices is discussed
within the framework of the extension of the standard model by a U(1)
symmetry.
This simple extension can lead to Solutions 1, 2 and 4 of ref.\cite{RRR}.

The selection rules obtained in the previous section could lead to other
types of mass matrices.
Here we discuss two examples corresponding to Solutions 3 and 5 of
ref.\cite{RRR} for the up-sector quarks.

Now we consider the $Z_4$ orbifold model where all of the quarks belong to
$T_1$ and the Higgs field corresponds to $T_2$.
Suppose that a state of $T_2$ sector develops the VEV.
We can choose the space group elements of the states so that the (1,2),
(2,1), (2,3) and (3,2) elements are forbidden.
For example we assign all of the first and third generations of the quarks
to the same fixed point, which is different from the fixed point
corresponding to the second generation.
Then we have the mass matrix proportional to the following matrix:
$$
\pmatrix{
m_{11} & 0 & \varepsilon^4 \cr
0 & \varepsilon^4 & 0 \cr
\varepsilon^4 & 0 & 1 \cr
}.
\eqno(3.1) $$
The (3,3) element corresponds to the renormalizable $T_1^2T_2$-coupling.
The nonvanishing elements with $\varepsilon^4 $ are obtained through
the $T_1^2T_2^5$-coupling.
If $m_{11}$ is enough suppressed, this matrix corresponds to Solution 3.
However, we have $m_{11}=\varepsilon^4$ if we take into account only the
selection rules due to the $H$-momentum conservation, the $Z_4$ invariance
of oscillators and the space group invariance.
Following ref.\cite{IG}, we can suppress this element if we assume the extra
U(1) symmetry to allow the (3,3), (1,3), (2,2) and (3,1) elements as (3.1).
Suppose that the $i$-th generation of the left-handed and right-handed
quarks have the U(1) charges $\alpha_i$ and $\alpha'_i$, respectively.
They should satisfy $\alpha_h=-\alpha_3-\alpha'_3$ and
$\alpha_3-\alpha_1=\alpha'_3-\alpha'_1=4\beta$, where $\alpha_h$ and
$\beta$ denote the U(1) charges of the Higgs field and the field with the
VEV, respectively.
Then the U(1) charges of the first generation satisfy
$\alpha_1+\alpha'_1+\alpha_h=-8\beta$.
Using this extra U(1) symmetry, we obtain $m_{11}=\varepsilon^8$, which
is derived from the $T_1^2T_2^9$-coupling.

Next we consider the $Z_6$-I orbifold model where the Higgs field corresponds
to $T_2$.
We assign the second and third generations of the quarks to $T_1$ and $T_2$,
respectively.
In addition the first generation of the quarks are assigned to $T_3$ states
with the $\theta$-eigenvalue $\gamma=\exp [2\pi i/3]$.
Suppose that a state of $T_1$ and a state of $T_3$ with
$\gamma=\exp [2\pi i/3]$ develop VEVs.
Then we obtain the mass matrix proportional to the following matrix:
$$
\pmatrix{
\varepsilon^9 & \varepsilon^9 & \varepsilon^2 \cr
\varepsilon^9 & \varepsilon^2 & \varepsilon \cr
\varepsilon^2 & \varepsilon & 1 \cr
}.
\eqno(3.2) $$
The (3,3) element corresponds to the renormalizable $T_2^3$-coupling.
The (2,3) and (3,2) elements of this matrix are induced by the
$T_1^2T_2^2$-coupling.
The (2,2) element is obtained through the $T_1^4T_2$-coupling.
The (1,3) and (3,1) elements are derived from the $T_1T_2T_3^3$-coupling.
Note that nonvanishing nonrenormalizable couplings should include
$T_3^{3\ell}$ $(\ell \geq 0)$ because each $T_3$ state is assumed to have
$\gamma=\exp [2\pi i/3]$.
The other elements correspond to the $T_1^3T_3^9$-coupling or
the $T_1^9T_3^3$-coupling.
This mass matrix corresponds to Solution 5 in the approximation that
$\varepsilon^9 $ is neglected.
In this case we do not need an extra symmetry to derive this form.
We can choose the space group elements of the states so that the space
group invariance forbids the (2,3) and (3,2) elements.
In this case the matrix (3.2) corresponds to Solution 3.

So far we have discussed the case where the field with the VEV belongs to
the non-oscillated twisted sector.
We could expect that fields in the untwisted sector or the oscillated
twisted sector develop VEVs
\def\thefootnote{\fnsymbol{footnote}}
\footnote[2]{The appearance of oscillated states are restricted in some cases
\cite{KKO}.}.
In this case the $H$-momentum conservation and the $Z_N$ invariance allow
nonrenormalizable couplings with lower powers, e.g. $UTTT$.
These couplings are also useful to derive the realistic mass matrices.

World-sheet instantons induce another suppression factor as
exp$[-aT]$, where $T$ is the moduli parameter and $a$ is a constant
\cite{Hamidi,DFMS}.
This factor lead to other types of the hierarchy in the quark masses
\cite{hiera}.

At last we comment on the CP phase.
In the ten-dimensional superstring theories, the CP is a good symmetry
\cite{CP}.
In ref.\cite{KL}, it is shown that the CP is unbroken in the
orbifold models without background anti-symmetric tensors.
The presence of the anti-symmetric tensors \cite{anti} breaks the CP
symmetry in a sense of the world-sheet.
Actually Yukawa couplings can have complex phases.
For the 3-point couplings of the $Z_3$ orbifold models, the selection
rule due to the space group invariance is very restrictive.
If we fix space group elements of the two states, the other state to
couple them is unique.
Then we obtain the diagonal mass matrices when we switch off
nonrenormalizable couplings, i.e. $\varepsilon \rightarrow 0$.
Hence we can eliminate complex phases by rephasing fields, even though
the Yukawa couplings have complex phases.
For the other orbifold models, $Z_4$ and $Z_6$, the selection rules are
not so restrictive.
However we can always eliminate complex phases in the case with
$\varepsilon \rightarrow 0$.
Thus the CP is unbroken in the effective field theory, although it is
broken in the world-sheet.
The nonrenormalizable couplings could induce the nontrivial appearance of
complex phases into the mass matrices.
It is important to investigate which assignment of the quarks to the space
group elements results in the mass matrices with the physical CP phase or
not.

\newpage
{\large \bf 4. Conclusion }
\vspace{0.8 cm}

We have studied the nonrenormalizable couplings in the orbifold models and
derived the nontrivial selection rules.
Some nonrenormalizable couplings lead naturally to the couplings with
strongly suppressed factors.
We can use the selection rules of the orbifold models to derive
interesting mass matrices, e.g. Solutions 3 and 5 of ref.\cite{RRR} as well
as Solutions 1, 2 and 4.
These analyses constrain assignments of the matter fields to the untwisted
and twisted sectors.
That is very useful for model building.
It is also important to investigate in which case the physical CP phase
appears.

In this paper we have restricted ourselves to the $Z_N$ orbifolds which are
obtained by products of three two-dimensional orbifolds.
We can discuss similarly for the other $Z_N$ orbifold models and the
$Z_M \times Z_N$ orbifold models \cite{ZMN}.

\vspace{0.8 cm}
\leftline{\large \bf Acknowledgement}
\vspace{0.8 cm}

The author would like to thank C.S.~Lim, H.~Nakano and S.~Stieberger for useful
discussions.



\newpage
\pagestyle{empty}
{\large Table 1. $H$-momenta}\\
For massless bosons, $H$-momenta of $T_k$ sectors are listed.
The elements  corresponding to the four-dimensional space-time are omitted.

\begin{tabular}{|c|c|c|c|c|c|}\hline
Orbifold & $v^i$ & $T_1$ & $T_2$ & $T_3$ & $T_4$ \\ \hline \hline
$Z_3$ & $(1,1,-2)/3$ & (1,1,1)/3 & & & \\
$Z_4$ & $(1,1,-2)/4$ & (1,1,2)/3 & (2,2,0)/4 & & \\
$Z_6$-I & $(-2,1,1)/6$ & (4,1,1)/6 & (2,2,2)/6 &(0,3,3)/6 & \\
$Z_6$-II & $(2,1,-3)/6$ & (2,1,3)/6 & (4,2,0)/6 &(0,3,3)/6 & (2,4,0)/6\\
\hline
\end{tabular}

\vspace{0.8 cm}
{\large Table 2. Allowed nonrenormalizable couplings}\\

\begin{tabular}{|c|c|c|}\hline
Orbifold & Coupling & Condition \\ \hline \hline
$Z_3$    & $T_1^{9\ell+3}$ & \\ \hline
         & $(T_1^2T_2)^{2\ell+1}$ & \\
$Z_4$    & $(T_1^2T_2)^{\ell+1}T_1^{4m}$ & $\ell+m=2n$ \\
         & $(T_1^2T_2)^{\ell+1}T_2^{2m}$ & $\ell+m=2n$ \\ \hline
         & $T_1^{36\ell+6}$ & \\
         & $T_2^{9\ell+3}$ & \\
$Z_6$-I  & $T_1^{2\ell}T_2^m$ & $\ell+m=9n+3, \ell >0, m >0$ \\
         & $T_1^{3\ell}T_3^m$ & $\ell+m=4n+2, \ell >0, m >0$ \\
         & $T_2^{3\ell+3}T_3^{2m}$ & $m > 0$ \\
         & $T_1^\ell T_2^m T_3^n$ & $\ell >0,m>0,n>0$, PGI \\ \hline
         & $T_1^{2\ell}T_2^m$ & $\ell+m=9n+3, \ell >0, m>0$ \\
         & $T_1^{6\ell}T_3^{2m}$ & $\ell+m=2n+1, \ell >0, m>0$ \\
         & $T_1^{2\ell}T_4^m$ & $\ell +2m=9n+3, 2\ell +m=9p+3,
                                \ell>0, m>0$ \\
$Z_6$-II & $T_2^{3\ell}T_3^{2m}$ & $\ell >0, m >0$ \\
         & $T_4^{3\ell}T_3^{2m}$ & $\ell >0, m >0$ \\
         & $T_1^{2\ell}T_2^mT_4^n$ & $\ell =2p+1,2\ell +2m+n=3q,
                                    \ell >0, m>0, n>0$, PGI \\
         & $T_1^\ell T_2^mT_3^nT_4^p$ & $\ell >0, n>0,
                                    m \ {\rm or } \ p>0 $, PGI \\
\hline
\end{tabular}

\begin{thebibliography}{99}

\bibitem{anom}
M.~Dine, N.~Seiberg and E.~Witten, Nucl.~Phys. {\bf B289} (1987) 585.

J.~Atick, L.~Dixon and A.~Sen, Nucl.~Phys. {\bf B292} (1987) 109.

M.~Dine, I.~Ichinose and N.~Seiberg, Nucl.~Phys. {\bf B293} (1987) 253.

\bibitem{anom1}
A.~Font, L.E.~Ib\'a\~nez, H.P.~Nilles and F.~Quevedo, Nucl.~Phys.
{\bf B307} (1988) 109; Phys.~Lett. {\bf B210} (1988) 101.

A.~Font, L.E.~Ib\'a\~nez, F.~Quevedo and A.~Sierra, Nucl.~Phys.
{\bf B331} (1990) 421.

\bibitem{anom2}
J.A.~Casas, E.K.~Katehou and C.~Mu\~noz, Nucl.~Phys. {\bf B317} (1989) 171.

J.A.~Casas and C.~Mu\~noz, Phys.~Lett. {\bf B209} (1988) 214; Phys.~Lett.
{\bf B214} (1988) 63.

\bibitem{mass}

C.D.~Froggatt and H.B.~Nielsen, Nucl.~Phys. {\bf B147} (1979) 277.

S.~Dimopoulos, Phys.~Lett. {\bf B129} (1983) 417.

M.~Leurer, Y.~Nir and N.~Seiberg, Nucl.~Phys. {\bf B398} (1993) 319;
Nucl.~Phys. {\bf B420} (1994) 468.

Y.~Nir and N.~Seiberg, Phys.~Lett. {\bf B309} (1993) 337.

\bibitem{IG}

L.E.~Ib\'a\~nez and G.G.~Ross, Phys.~Lett. {\bf B332} (1994) 100.

\bibitem{mu}
J.E.~Kim and H.P.~Nilles, Phys.~Lett. {\bf B138} (1984) 150; Phys.~Lett.
{\bf B263} (1991) 79; Mod.~Phys.~Lett. {\bf A9} (1994) 3575.

E.J.~Chun, J.E.~Kim and H.P.~Nilles, Nucl.~Phys. {\bf B370} (1992) 105.

J.A.~Casas and C.~Mu\~noz, Phys.~Lett. {\bf B306} (1993) 288.

\bibitem{neutrino}
M.~Masip, Phys.~Rev. {\bf D46} (1992) 3601;
Phys.~Rev. {\bf D47} (1993) 3071.

N.~Haba, C.~Hattori, M.~Matsuda, T.~Matsuoka and D.~Mochinaga, Phys.~Lett.
{\bf B337} (1994) 63; Prog.~Theor.~Phys. {\bf 92} (1994) 153.

\bibitem{Orbi}
L.~Dixon, J.~Harvey, C.~Vafa and E.~Witten, Nucl.~Phys. {\bf B261} (1985)
 678; Nucl.~Phys. {\bf B274} (1986) 285.

L.E.~Ib\'a\~nez, J.~Mas, H.P.~Nilles and F.~Quevedo, Nucl.~Phys. {\bf B301}
 (1988) 157.

Y.~Katsuki, Y.~Kawamura, T.~Kobayashi, N.~Ohtsubo, Y.~Ono and \\
K.~Tanioka, Nucl.~Phys. {\bf B341} (1990) 611.

\bibitem{Hamidi}
S.~Hamidi and C.~Vafa, Nucl.~Phys. {\bf B279} (1987) 465.

\bibitem{DFMS}
L.~Dixon, D.~Friedan, E.~Martinec and S.~Shenker, Ncul.~Phys.
{\bf B282} (1987) 13.

\bibitem{KO1}
T.~Kobayashi and N.~Ohtsubo, Phys.~Lett. {\bf B245} (1990) 441.

\bibitem{KO2}
T.~Kobayashi and N.~Ohtsubo, Int.~J.~Mod.~Phys. {\bf A9} (1994) 87.

\bibitem{CM}
J.A.~Casas, F.~Gomez and C.~Mu\~noz, Int.~J.~Mod.~Phys. {\bf A8} (1993) 455.

\bibitem{Cvetic}
M.~Cveti\u{c}, Phys.~Rev.~Lett. {\bf 59} (1987) 1795;
Phys.~Rev.~Lett. {\bf 59} (1987) 2829.

\bibitem{Nonreno}
A.~Font, L.E.~Ib\'a\~nez, H.P.~Nilles and F.~Quevedo, Phys.~Lett.
{\bf B213} (1988) 274.

\bibitem{FMS}
D.~Friedan, E.~Martinec and S.~Shenker, Ncul.~Phys. {\bf B271} (1986) 93.

\bibitem{RRR}
R.~Ramond, R.G.~Roberts and G.G.~Ross, Nucl.~Phys. {\bf B406} (1993) 19.

\bibitem{KKO}
H.~Kawabe, T.~Kobayashi and N.~Ohtsubo, Phys.~Lett. {\bf B325} (1994) 77;
Nucl.~Phys. {\bf B434} (1995) 210.

T.~Kobayashi, preprint Kanazawa-94-10 (hep-ph/9406238)
to be published in Int.~J.~Mod.~Phys. A.

\bibitem{hiera}
L.E.~Ib\'a\~nez, Phys.~Lett. {\bf B181} (1986) 269.

J.A~Casas and C.~Mu\~noz, Nucl.~Phys. {\bf B332} (1990) 189.

J.A~Casas, F.~Gomez and C.~Mu\~noz, Phys.~Lett. {\bf B292} (1992) 42.

\bibitem{CP}
C.S.~Lim, Phys.~Lett. {\bf B256} (1991) 233.

M.~Dine, R.G.~Leigh and D.A.~MacIntire, Phys.~Rev.~Lett.
{\bf 69} (1992) 2030.

K.~Choi, D.B.~Kaplan and A.E.~Nelson, Nucl.~Phys. {\bf B391} (1993) 515.

\bibitem{KL}
T.~Kobayashi and C.S.~Lim, Phys.~Lett. {\bf B343} (1995) 122.

\bibitem{anti}
J.~Erler, D.~Jungnickel and J.~Lauer, Phys.~Rev. {\bf D45} (1992) 3651.

D.~Jungnickel, J.~Lauer, M.~Spali\'nski and S.~Stieberger,
Mod.~Phys.~Lett. {\bf A7} (1992) 3059.

J.~Erler, D.~Jungnickel, M.~Spali\'nski and S.~Stieberger,
Nucl.~Phys. {\bf B397} (1993) 379.

M.~Sakamoto, Nucl.~Phys. {\bf B414} (1994) 267.

\bibitem{ZMN}
A.~Font, L.E.~Ib\'a\~nez and F.~Quevedo, Phys.~Lett. {\bf B217} (1989) 272.

T.~Kobayashi and N.~Ohtsubo, Phys.~Lett. {\bf B262} (1991) 425.



\end{thebibliography}
\end{document}